\pdfoutput=1
\documentclass[aps,pra,reprint,superscriptaddress,nofootinbib,floatfix,
  showkeys]{revtex4-2}

\usepackage{amsmath,amssymb,bm}
\usepackage{booktabs}
\usepackage{graphicx}
\usepackage{microtype}
\usepackage{xcolor}
\usepackage[colorlinks=true,linkcolor=blue,citecolor=blue,urlcolor=blue]{hyperref}
\hypersetup{
  pdftitle={Variance-weighted measurement planning and rank-aware adaptive precision for quantum subspace eigensolvers},
  pdfauthor={Ginanjar Utama and Hermawan Kresno Dipojono},
  pdfsubject={Preregistered end-to-end study of weighted k-commuting grouping, paid-pilot Ritz replanning, rank-aware adaptive precision stopping, and noise-gated readout circuits}
}

\newcommand{\ket}[1]{\lvert #1\rangle}
\newcommand{\bra}[1]{\langle #1\rvert}
\newcommand{\avg}[1]{\langle #1\rangle}
\newcommand{\mHa}{\mathrm{mHa}}

\newcommand{\pkg}{\texttt{clifford\_qc}}
\newcommand{\Rhat}{\hat{R}}
\newcommand{\Remp}{R_{\mathrm{emp}}}
\newcommand{\CZmax}{\mathrm{CZ}_{\max}}
\newcommand{\tightcolumns}{\setlength{\tabcolsep}{3.5pt}}
\newcommand{\compactcolumns}{\setlength{\tabcolsep}{2.4pt}}
\usepackage{catchfile}
\makeatletter
\newcommand{\tableinput}[1]{%
  \noalign{\CatchFileDef{\wp@caughtrows}{#1}{}%
    \global\let\wp@tablerows\wp@caughtrows}%
  \wp@tablerows}
\makeatother

\newcommand{\wpBehTwoRitzPilot}{19,992}
\newcommand{\wpBehTwoRitzRmse}{1.260}
\newcommand{\wpBiasRedMax}{13.4}
\newcommand{\wpBiasRedMin}{5.2}
\newcommand{\wpBootAnalysisReps}{2,000}
\newcommand{\wpBootAnalysisSeed}{2026112399}
\newcommand{\wpBootSolves}{49,696}
\newcommand{\wpExhausted}{4}
\newcommand{\wpFailed}{0}
\newcommand{\wpLihLdfFcLookTwoHi}{3.809}
\newcommand{\wpLihLdfFcLookTwoLo}{1.735}
\newcommand{\wpLihLdfFcLookTwoSE}{2.064}
\newcommand{\wpLihLdfFcLookZeroSE}{35.68}
\newcommand{\wpLihRitzFcRRatio}{1.628}
\newcommand{\wpLihSiFcLookTwoHi}{2.027}
\newcommand{\wpLihSiFcLookTwoLo}{1.958}
\newcommand{\wpLihSiFcLookTwoSE}{1.996}
\newcommand{\wpLihSiFcLookZeroSE}{13.88}
\newcommand{\wpLihSiFcRRatio}{1.397}
\newcommand{\wpLihSiQwcLookTwoSE}{1.539}
\newcommand{\wpLihSiQwcRRatio}{3.799}
\newcommand{\wpMeasuredSolves}{6,272}
\newcommand{\wpNoisyAnchorCount}{36}
\newcommand{\wpNumpyVer}{2.5.2}
\newcommand{\wpPartAAcq}{832}
\newcommand{\wpPartBAcq}{576}
\newcommand{\wpPilotShotsExact}{5,914,592}
\newcommand{\wpPoneaMedMax}{0.400}
\newcommand{\wpPoneaMedMin}{0.064}
\newcommand{\wpPoneaPassCount}{8}
\newcommand{\wpPoneaSpeedupMax}{13.6}
\newcommand{\wpPoneaSpeedupMin}{2.5}
\newcommand{\wpPoneaTotalCount}{9}
\newcommand{\wpPthreePassCount}{6}
\newcommand{\wpPthreeTotalCount}{6}
\newcommand{\wpPtwoHighTightBiasMax}{2.97}
\newcommand{\wpPtwoHighTightBiasMin}{1.17}
\newcommand{\wpPtwoHighUncBiasMax}{15.55}
\newcommand{\wpPtwoHighUncBiasMin}{13.82}
\newcommand{\wpPtwoLowTightFalseMax}{4}
\newcommand{\wpPtwoLowTightFalseMin}{1}
\newcommand{\wpPtwoLowUncFalseMax}{14}
\newcommand{\wpPtwoLowUncFalseMin}{13}
\newcommand{\wpPythonVer}{3.12.14}
\newcommand{\wpRRatioMax}{3.799}
\newcommand{\wpRRatioMin}{0.171}
\newcommand{\wpRmseRedMax}{7.1}
\newcommand{\wpRmseRedMin}{2.4}
\newcommand{\wpScFullBiasMax}{134.9}
\newcommand{\wpScFullBiasMin}{61.8}
\newcommand{\wpScipyVer}{1.18.0}
\newcommand{\wpSeedRoot}{2026112301}
\newcommand{\wpStretchedLdfQwcMeanM}{4.883}
\newcommand{\wpStretchedLdfRmse}{0.809}
\newcommand{\wpStretchedRitzRatioLdf}{0.820}
\newcommand{\wpStretchedRitzRmse}{0.467}
\newcommand{\wpStretchedRitzWidth}{2.68}
\newcommand{\wpStretchedSiKTwoMeanM}{3.418}
\newcommand{\wpStretchedSiKTwoSavePct}{30.0}
\newcommand{\wpStretchedSiKTwoVsSiQwcPct}{14.6}
\newcommand{\wpStretchedWords}{10,341}
\newcommand{\wpTargetMet}{1,404}
\newcommand{\wpTotalAcq}{1,408}
\newcommand{\wpTotalShotsB}{3.14}
\newcommand{\wpTotalShotsExact}{3,144,593,750}
\newcommand{\wpWidthBootRatioMax}{2.88}
\newcommand{\wpWidthBootRatioMin}{2.78}
\newcommand{\wpWidthExhErrMax}{0.694}
\newcommand{\wpWidthExhErrMin}{0.023}
\newcommand{\wpWidthMeanRatioMax}{5.78}
\newcommand{\wpWidthMeanRatioMin}{5.43}
\newcommand{\wpWidthRmseMax}{0.53}
\newcommand{\wpWidthRmseMin}{0.45}

\begin{document}

\title{Variance-weighted measurement planning and rank-aware adaptive precision for quantum subspace eigensolvers}

\author{Ginanjar Utama}
\email{ginanjar.utama@gmail.com}
\affiliation{Department of Engineering Physics, Institut Teknologi Bandung,
Bandung, Indonesia}

\author{Hermawan Kresno Dipojono}
\email{dipojono@itb.ac.id}
\affiliation{Department of Engineering Physics, Institut Teknologi Bandung,
Bandung, Indonesia}

\date{November 29, 2026}

\begin{abstract}
Quantum subspace eigensolvers must control the joint sampling uncertainty of
the projected Hamiltonian and overlap, not merely the number of measurement
settings. We integrate variance-weighted commuting-Pauli grouping, independent
paid-pilot Ritz replanning, and per-setting entangling-gate limits with a
density-free, rank-aware adaptive precision controller. A preregistered study
comprises \wpTotalAcq{} acquisitions and \wpTotalShotsB{} billion simulated
physical shots on fixed eight- and ten-qubit molecular banks. Weighted sorted
insertion meets the declared shot-cost and reliability criteria in
\wpPoneaPassCount{}/\wpPoneaTotalCount{} matched-block comparisons, requiring
\wpPoneaSpeedupMin{}--\wpPoneaSpeedupMax{} times fewer mean shots in the
passing cells; the remaining cell ties on the discrete acquisition grid, so
the global criterion fails. Paid-pilot replanning helps BeH\(_2\), but on LiH
qubit-wise commuting (QWC) readout is cheaper than fully commuting readout,
illustrating the limitations of state-independent variance proxies. Both
weighted two-qubit-block arms meet the rank-stress criterion on stretched LiH,
with no reduced-rank stops. An optional bootstrap-width guard instead incurs
\wpWidthMeanRatioMin{}--\wpWidthMeanRatioMax{} times the mean shots, including
\wpExhausted{} exhaustions. Tight noise gating meets all
\wpPthreePassCount{}/\wpPthreeTotalCount{} relative two-qubit-noise criteria,
but does not ensure absolute accuracy: the combined readout-noise regime
produces \wpScFullBiasMin{}--\wpScFullBiasMax{}~mHa systematic bias in all
arms, including QWC. These are simulations with fixed-bank energy targets,
not hardware demonstrations or quantum advantage claims.
\end{abstract}


\maketitle

\section{Introduction}
\label{sec:introduction}

Quantum subspace diagonalization (QSD), quantum subspace expansion (QSE), and
operator-generated subspace methods reconstruct a small generalized
eigenproblem \(H_{\mathrm{sub}} c = E S c\) from measured matrix elements on
one or more prepared states
\cite{mcclean2017qse,huggins2020nonorthogonal,stair2020krylov,utama2026operatorcentric,utama2026acase}.
Because finite-shot perturbations in \(S\) are amplified near its small
singular values \cite{epperly2022theory,lee2024sampling,zhang2024measurement},
practical accuracy depends simultaneously on two coupled layers:
\begin{enumerate}
  \item \textbf{Measurement planning:} how the Pauli words required to assemble
    \(S\) and \(H_{\mathrm{sub}}\) are partitioned into commuting measurement
    settings, compiled into readout circuits, and bounded against two-qubit
    entangling-gate noise
    \cite{verteletskyi2020grouping,izmaylov2020unitary,crawford2021efficient,miller2024hardware,dalfavero2025kcommutativity,burns2024galic}.
  \item \textbf{Adaptive precision stopping:} how physical shots are allocated
    across settings over sequential looks, when the solver stops, and how
    overlap-rank instability or sampling uncertainty is guarded
    \cite{efron1979bootstrap,utama2026cliffordqc}.
\end{enumerate}

Previous work already establishes the benefits of weighted grouping,
covariance-aware measurement allocation, and hardware-aware commutation
\cite{crawford2021efficient,yen2023covariance,dalfavero2025kcommutativity,burns2024galic}.
In a subspace eigensolver, however, optimizing a single Hamiltonian expectation
or counting settings does not determine the cost of the generalized eigenvalue
estimate: its sensitivities change with the measured Ritz vector, the overlap
may lose rank, and entangling readout errors can distort both matrices. Neither
an unweighted graph-coloring heuristic nor a state-independent shot proxy can
capture all three effects. In particular, reducing sampling variance alone does
not control bias under noisy readout.

Our contribution is the integration and preregistered evaluation of these
ingredients in the density-free \pkg{} workflow \cite{utama2026cliffordqc},
rather than a new commutation relation or a new noise-mitigation theorem. We
use a common fixed-bank generalized eigenproblem, a physical-shot budget
including every replanning pilot, and a shared rank-aware stopping controller
to compare QWC, block-commuting, and fully commuting (FC) plans. The
gate-level simulator supplies an independent infinite-shot noisy energy
diagnostic, while the controller receives only joint bitstring histograms.
The design isolates the impact of changing the grouping policy at fixed block
size, tests transfer to an engineered near-cutoff overlap, and separates
relative improvements under two-qubit noise from absolute energy accuracy
under combined readout noise.

\section{Methods}
\label{sec:methods}

\subsection{Density-free graph-state measurement planning}
\label{sec:methods_planning}

Given a fixed single-reference generator bank \(\{A_i\}_{i=1}^{m}\) and qubit
Hamiltonian \(H = \sum_P h_P P\), the projected overlap and Hamiltonian
matrices have entries \(S_{ij} = \bra{\psi} A_i^\dagger A_j \ket{\psi}\) and
\(H_{ij} = \bra{\psi} A_i^\dagger H A_j \ket{\psi}\).  Expanding every product
in the Pauli basis yields a single shared word bank \(\mathcal{W}\) and linear
functionals \(\{f_{ij}^{(S)}, f_{ij}^{(H)}\}\) stored in packed coefficient
form \cite{utama2026cliffordqc}.  Crucially, no \(2^n \times 2^n\) density
matrix is ever formed by the consumer: the measurement planner and adaptive
controller interact with the reference solely through compiled measurement
settings and sampled bitstring histograms.

Each commuting group \(G_g \subseteq \mathcal{W}\) is constrained by
\(k\)-commutativity \cite{dalfavero2025kcommutativity}: the \(n\)-qubit
register is partitioned into fixed contiguous blocks of size at most \(k\),
and every pair of words must commute on each block (\(k=1\) gives QWC,
\(k=2\) restricts entangling readout circuits to disjoint qubit pairs of depth
at most~1, and \(k=n\) gives unconstrained FC).  Every group is compiled into a
Clifford readout circuit using the NumPy-only graph-state synthesizer
(\texttt{method="graph\_state"}), with each Pauli word assigned to exactly one
setting (\texttt{assigned} pooling).

\subsection{Variance-weighted sorted insertion and Ritz replanning}
\label{sec:methods_weighting}

To prioritize shot variance rather than clique count, we replace unweighted
largest-degree-first (LDF) coloring with weighted sorted insertion (SI)
\cite{crawford2021efficient}.  For a nonnegative word-weight map
\(w: \mathcal{W} \to \mathbb{R}_{\ge 0}\), words are ordered by descending
\(w(P)\) and inserted greedily into the first existing \(k\)-commuting group
compatible with every word already in that group.

We evaluate two density-free weight definitions:
\begin{enumerate}
  \item \textbf{Bank-weighted SI (\texttt{si\_*}):} uses only the static
    operator coefficients across the projected matrix functionals,
    \(w_{\mathrm{bank}}(P) = \sqrt{\sum_{\ell} |c_{\ell, P}|^2}\), requiring
    zero quantum shots.
  \item \textbf{Two-stage paid-pilot Ritz replanning (\texttt{ritz\_si\_*}):}
    first spends an independent pilot budget on a bank-weighted plan to
    estimate a preliminary Ritz eigenvector \(\hat{c}\) and weakest overlap
    eigenvector \(\hat{u}\), then computes first-order sensitivity weights
    combining the Ritz functional \(g_{\mathrm{Ritz}}(\hat{c})\) and normalized
    overlap-margin functional \(g_{\mathrm{ov}}(\hat{u})\) with mixing weight
    \(\lambda_{\mathrm{ov}} = 0.5\), replans the \(k\)-commuting partition
    once, and runs adaptive acquisition on the replanned settings.  To prevent
    cost hiding, every pilot shot is billed unconditionally to
    \texttt{total\_shots} and subtracted from the remaining look targets so
    single-stage and two-stage arms share the same cumulative physical-shot
    grid.
\end{enumerate}

To connect the grouping objective to the generalized eigenproblem, let
\(c^\dagger S c=1\) for a simple, retained Ritz root. Its first-order
perturbation is
\begin{equation}
  \mathrm{d}E=c^\dagger(\mathrm{d}H_{\rm sub}-E\,\mathrm{d}S)c
             =\sum_{P\ne I}a_P\,\mathrm{d}\avg{P}.
  \label{eq:ritz_derivative}
\end{equation}
The replanning weights are \(\sqrt{a_P^2+(\lambda_{\rm ov}b_P)^2}\), where
\(b_P\) is the coefficient of the weakest normalized overlap-margin
derivative, when available. Both derivatives are evaluated on pilot data.
This linearization is local to a stable retained frame, motivating the
separate rank guard below.

For a fixed real linear functional \(F=\sum_{P\ne I}a_P P\), define the
single-shot variance \(v_g=\operatorname{Var}(\sum_{P\in G_g}a_P X_P)\), with
\(X_P\in\{-1,1\}\) the jointly measured signed parity. Thus
\(v_g=\bm a_g^\mathsf{T}\Sigma_g\bm a_g\), including all within-setting
covariances. With independent setting samples and \(N=\sum_g N_g\), the
variance is \(\sum_g v_g/N_g\), minimized in the continuous allocation
relaxation by \(N_g\propto\sqrt{v_g}\). The minimum is
\((\sum_g\sqrt{v_g})^2/N\). Comparing to optimal one-word-per-setting
allocation gives the state-independent and empirical optimal-shot ratios
\cite{crawford2021efficient,yen2023covariance}:
\begin{equation}
  \begin{aligned}
  \Rhat &= \frac{(\sum_P |a_P|)^2}
                 {(\sum_g\sqrt{\sum_{P\in G_g}a_P^2})^2},\\
  \Remp &= \frac{(\sum_P |a_P|\sqrt{\hat v_P})^2}
                 {(\sum_g\sqrt{\hat v_g})^2},
  \end{aligned}
  \label{eq:rhat_remp}
\end{equation}
where \(\hat v_P\) and \(\hat v_g\) are the estimated marginal and grouped
variances, respectively, with the implementation's small-sample variance
floor. The proxy \(\Rhat\) replaces all marginal variances by one and
off-diagonal covariances by zero; it is not an upper bound on the actual
shot ratio. For the static bank proxy, replace \(|a_P|\) by
\(w_{\rm bank}(P)\), which aggregates multiple matrix functionals. Larger
ratios indicate a greater potential saving relative to the corresponding
ungrouped allocation, not an absolute prediction of adaptive shot cost.
Since both marginal variances and covariances change relative to the proxy,
\(\Remp/\Rhat\) is a calibration diagnostic, not an isolated measure of
covariance. Terminal Ritz metrics use the estimated root and are descriptive,
not an oracle available to the initial planner.

\subsection{GALIC-style two-qubit noise gating}
\label{sec:methods_galic}

Following the hardware-aware bias bound of GALIC \cite{burns2024galic}, our
\texttt{NoiseGatePolicy} enforces an explicit per-setting two-qubit gate cap
during greedy sorted insertion.  Given a native two-qubit depolarizing error
\(p_{2\mathrm{q}}\), routing overhead multiplier \(r_{\mathrm{route}}\), and
relative bias tolerance \(\varepsilon_{\mathrm{bias}}\), a candidate word
insertion into group \(G_g\) is accepted only if the compiled graph-state
readout circuit for \(G_g \cup \{P\}\) satisfies
\begin{equation}
  N_{\mathrm{CZ}}^{\mathrm{log}}(G_g \cup \{P\})
  \le \left\lfloor \frac{\ln(1 - \varepsilon_{\mathrm{bias}})}{r_{\mathrm{route}} \ln(1 - p_{2\mathrm{q}})} \right\rfloor.
  \label{eq:galic_cap}
\end{equation}
If no existing commuting group can absorb \(P\) within the cap, \(P\) opens a
new setting (always admissible with \(0\) two-qubit gates).  In our study, tight
gating (\(\varepsilon_{\mathrm{bias}} = 0.01\) at \(p_{2\mathrm{q}}=0.002\) or
\(0.05\) at \(p_{2\mathrm{q}}=0.010\), with \(r_{\mathrm{route}}=3\)) yields
\(\CZmax = 1\), while moderate gating yields \(\CZmax = 3\). This is a
heuristic survival-probability gate budget (the implementation uses the
ceiling of routed gate count). It is not a bound on the absolute error of a
nonlinear generalized eigenvalue, and it constrains neither one-qubit nor
readout error.

\subsection{Rank-aware adaptive precision stopping}
\label{sec:methods_stopping}

All arms are evaluated inside the unified raw-Ritz \texttt{sampled\_precision}
controller \cite{utama2026cliffordqc}.  At each predeclared look \(t \in
\{0,\dots,L-1\}\) along a matched total physical-shot schedule, the controller
assembles \((\hat{S}_t, \hat{H}_t)\), solves the thresholded generalized
eigenproblem with declared cutoff \(\tau_S\) and condition cap, estimates the
delta-method Ritz standard error \(\widehat{\mathrm{SE}}_t\), and evaluates
\texttt{RankStopPolicy} (two consecutive looks, margin error level
\(\delta_r=0.05\)). With \(D=\operatorname{diag}(\hat S_t)\), let
\(\lambda_-\) and \(\lambda_+\) denote the smallest and largest eigenvalues of
the normalized overlap \(D^{-1/2}\hat S_t D^{-1/2}\) on the live directions.
For full-rank eligibility the three estimated margins are
\begin{equation}
  \begin{aligned}
  m_1 &= \lambda_- - \tau_S,\\
  m_2 &= \lambda_- - \tau_{\rm rel}\lambda_+,\\
  m_3 &= \lambda_- - \lambda_+/\kappa_{\max}.
  \end{aligned}
  \label{eq:rank_margins}
\end{equation}
Their lower limits are \(m_j-z_{1-\delta_r/(6L)}\,\widehat{\mathrm{SE}}(m_j)\),
using the measured normalized-eigenvalue derivatives (including the
diagonal-normalization terms and covariances between modes). Whenever rank
is unresolved, the controller splits the next round budget between
margin-targeted overlap shots (half of the round if \(\widehat{\mathrm{SE}}_t >
\tau_{\mathrm{SE}}\), all of the round otherwise) and Ritz-variance-targeted
shots.  Stopping requires \(\widehat{\mathrm{SE}}_t \le 1.6\,\mHa\) and full
declared rank with positive lower margins on two consecutive looks.

The controller resamples joint outcome histograms within settings and
re-solves the thresholded pencil \cite{efron1979bootstrap}. Let \(q_p\) be
the \(p\)-quantile of the bootstrap Ritz estimates and
\(s_b=(q_{0.84}-q_{0.16})/2\). The reported bootstrap-spread interval is
centered on the measured estimate \(\hat E_t\), not on the bootstrap
quantiles:
\begin{equation}
  \hat E_t\pm z_{1-\delta/(2L)}\,s_b.
  \label{eq:bootspread}
\end{equation}
The delta-method interval uses the same multiplier with
\(\widehat{\mathrm{SE}}_t\) in place of \(s_b\). We also test an opt-in
width guard (\texttt{si\_fc\_width}) requiring the half-width of
Eq.~\eqref{eq:bootspread} at nominal 95\% to be at most \(1.6\,\mHa\).
These are asymptotic and heuristic intervals: the look adjustment does not
prove finite-sample or anytime-valid coverage under adaptive allocation and
rank truncation.

\subsection{Gate-level noisy readout simulator}
\label{sec:methods_noisy_sim}

The sampling device propagates an ideal prepared state through each compiled
Clifford readout circuit. A \(d\)-dimensional depolarizing channel on the
gate's support \(Q\) acts as
\(\mathcal D_{Q,\lambda}(\rho)=(1-\lambda)\rho+\lambda(I_Q/d)\otimes
\operatorname{Tr}_Q\rho\), with the tensor factors placed on their original
qubits. In the simulator, \(p_{1\mathrm{q}}\) and \(p_{2\mathrm{q}}\) are
total nonidentity Pauli-error probabilities, so the strengths after each
local gate and logical CZ are, respectively,
\begin{equation}
  \begin{aligned}
  \lambda_1 &= \tfrac43p_{1\mathrm{q}},\\
  \lambda_2 &= 1-\left(1-\tfrac{16}{15}p_{2\mathrm{q}}\right)^{r_{\rm route}}.
  \end{aligned}
  \label{eq:depolarizing}
\end{equation}
Independent symmetric classical bit flips with probability \(p_{\mathrm{ro}}\)
are then applied to the outcome distribution. The gate-cap heuristic in
Eq.~\eqref{eq:galic_cap} uses \(p_{2\mathrm{q}}\), not \(\lambda_2\).
Ideal settings use a statevector, while noisy gate evolution uses a density
matrix \emph{only inside the simulated device}. The density-free claim
applies to the coefficient planner, reconstruction and controller, not to the
classical noise simulator. Physical shots are multinomial draws from the
resulting joint distribution; routed noise is modeled by a uniform multiplier,
not by explicit hardware routing or calibrated preparation noise.

Exact noisy word expectations from the same compiled settings define
\(E_\infty^{\rm noisy}\) using the same numerical cutoff. Its difference
from \(E_{\rm exact}\) diagnoses systematic readout-circuit bias. Separately,
for terminal finite-shot errors \(e_i=\hat E_i-E_{\rm exact}\), we report
\(\overline{e^2}=\bar e^2+\overline{(e_i-\bar e)^2}\); the bias-squared
share is \(\bar e^2/\overline{e^2}\). The finite-sample empirical bias
\(\bar e\) need not equal the infinite-shot bias because eigenvalue estimation
and stopping are nonlinear.

\section{Preregistered Experimental Protocol}
\label{sec:protocol}

The complete protocol was frozen prior to evaluation in the result-free
declaration \path{docs/restart/weighted_precision_declaration.json},
binding the SHA-256 digests of every
tracked Python file under \texttt{clifford\_qc/} and \texttt{benchmarks/}, all
four FCIDUMP files \cite{knowles1989fcidump}, historical anchor records, and
the illustrative superconducting device card.  Development seed~\texttt{501}
was used solely to verify bank anchors and calibrate shot schedules prior to
freezing and is excluded from evaluation; evaluation seeds derive
deterministically from \texttt{seed\_root = \wpSeedRoot} via NumPy
\texttt{SeedSequence} with zero collisions across all \(2{,}928\) streams.

\begin{table*}[t]
\centering
\caption{Molecular test banks, exact projected energy targets, minimum
normalized overlap eigenvalue and cutoff (shown as \(\lambda_{\min}; \tau_S\)),
Pauli word counts, and static
Part~A compiled \texttt{graph\_state} plan metrics.  All numbers are generated
from the frozen evaluation record.}
\label{tab:protocol_banks}
\tightcolumns
\small
\begin{tabular}{l r l r l r r r r}
\toprule
Bank & Words & Rank (\(\lambda_{\min}; \tau_S\)) & Exact \(E\) (Ha) & Arm & Settings & \(\Rhat\) & \(\mathrm{CZ}_{\mathrm{tot}}\) (\(\CZmax\)) & \(d_{2\mathrm{q},\max}\) \\
\midrule
\tableinput{tables/protocol_banks.tex}
\bottomrule
\end{tabular}
\end{table*}

Table~\ref{tab:protocol_banks} summarizes the four molecular banks and their
static compiled plans. The reference is
\(\cos(0.3)\ket{a}+e^{0.2i}\sin(0.3)\ket{b}\), where \(a,b\) are occupation
bitstrings (ordered by spatial orbital and spin). Define
\(G_{st}=T_{st}-T_{st}^\dagger\), where
\(T_{st}=a_{t\uparrow}^\dagger a_{t\downarrow}^\dagger
a_{s\downarrow}a_{s\uparrow}\) transfers an electron pair from spatial orbital
\(s\) to \(t\). The eight-qubit banks have four electrons,
\(a=11110000\), \(b=11001100\), and directions
\(I,(I+G_{12})/\sqrt2,(I+G_{13})/\sqrt2\). For stretched LiH, the core is
frozen and five active spatial orbitals contain two electrons; the bitstrings
are \(1100000000\) and \(0011000000\), and the directions are
\(I,I+0.5G_{01},G_{02},G_{03}\). The near-dependent direction is deliberately
engineered to test the controller, not discovered through adaptive ansatz
selection. In all cases the target is the lowest exact projected Ritz root
with the declared cutoffs, not the exact molecular ground energy.

The two study parts proceed as follows:
\begin{itemize}
  \item \textbf{Part~A (ideal shots, \wpPartAAcq{} acquisitions):} Three
    well-conditioned eight-qubit banks --- BeH\(_2\) (1,223 words), LiH (3,167
    words), and H\(_4\) (3,051 words), each rank~3 with \(\lambda_{\min} =
    0.2106 \gg \tau_S = 0.005\) --- tested across 8 arms
    (\texttt{ldf\_qwc}, \texttt{si\_qwc}, \texttt{ldf\_k2}, \texttt{si\_k2},
    \texttt{ldf\_fc}, \texttt{si\_fc}, \texttt{ritz\_si\_fc},
    \texttt{si\_fc\_width}), 32 outer replicas, 32 bootstrap replicates, and a
    six-look matched cumulative schedule (\([0.04, 0.10, 0.25, 0.625, 1.5625,
    3.90625]\times 10^6\) shots for BeH\(_2\)/LiH; up to \(9.766\times 10^6\)
    for H\(_4\)); plus the engineered near-cutoff ten-qubit stretched LiH bank
    (\wpStretchedWords{} words, rank~4 with \(\lambda_{\min} = 0.10417\)
    just above \(\tau_S = 0.104\)) across 4 arms (\texttt{ldf\_qwc},
    \texttt{si\_qwc}, \texttt{si\_k2}, \texttt{ritz\_si\_k2}) and 16 outer
    replicas, 16 bootstrap replicates, and cumulative targets of
    \([0.5,1.25,3.125,7.8125,19.53125]\times10^6\) shots.
  \item \textbf{Part~B (gate-level noisy readout, \wpPartBAcq{} acquisitions):}
    BeH\(_2\), LiH, and H\(_4\) (16 outer replicas each) across 4 arms
    (\texttt{si\_qwc}, \texttt{si\_fc\_unconstrained},
    \texttt{si\_fc\_galic\_tight}, \texttt{si\_fc\_galic\_moderate}) under
    three noise regimes: (i)~\texttt{depol2q\_p002\_r3} (\(p_{2\mathrm{q}} =
    0.002, r_{\mathrm{route}} = 3, p_{1\mathrm{q}}=p_{\mathrm{ro}}=0\));
    (ii)~\texttt{depol2q\_p010\_r3} (\(p_{2\mathrm{q}} = 0.010,
    r_{\mathrm{route}} = 3, p_{1\mathrm{q}}=p_{\mathrm{ro}}=0\)); and
    (iii)~\texttt{sc\_card\_full}
    (\(p_{1\mathrm{q}} = 0.001, p_{2\mathrm{q}} = 0.010, r_{\mathrm{route}} =
    3, p_{\mathrm{ro}} = 0.02\)).
\end{itemize}

For a plan with \(M\) settings and first-look target \(T_0\), the initial
uniform acquisition uses \(\max(2,\lfloor T_0/M\rfloor)\) shots per setting.
Later rounds use the remaining cumulative target, subject to the declared
minimum budget; thus the first look can differ by integer rounding across
arms, while subsequent looks share the same physical-shot grid here. The
\(2.5\times\) geometric look spacing was chosen deliberately to span up to
roughly two orders of magnitude in total physical shots across only \(L=5\)--\(6\)
sequential looks, trading coarser shot-count quantization against the
sequential multiple-look Bonferroni multipliers
(\(z_{1-\delta/(2L)}\) and \(z_{1-\delta_r/(6L)}\)) and per-look
resampling cost; a finer grid (such as \(1.5\times\) or \(\sqrt{2}\times\))
would resolve smaller cost differences at the expense of more looks, wider
schedule-adjusted intervals, and stricter rank-margin thresholds. The
two-stage arms pay a separate same-block-size pilot (nominally 20,000 shots
for 8q and 250,000 for 10q) and subtract its realized shots before acquiring
fresh data; pilot outcomes are not pooled into the final acquisition.

Four primary endpoints were frozen in \texttt{check\_weighted\_precision.py}
prior to execution:
\begin{itemize}
  \item \textbf{P1a:} On the nine eight-qubit matched-\(k\) pairs
    (\(k \in \{1,2,8\}\)), weighted SI passes a cell if median shot ratio
    \(\le 0.80\) (or strictly lower mean ratio on a median tie), \(\ge 30/32\)
    \texttt{target\_met} stops, \(0\) reduced-rank stops, and non-rejected
    unconditional 95\% bootstrap coverage (exact one-sided binomial lower tail
    \(p_{\mathrm{lo}} \ge 0.05/9\)).
  \item \textbf{P1b (descriptive):} Report matched-arm shot ratios, RMSE, 95\%
    coverage, and median \(\Remp/\Rhat\) for paid-pilot \texttt{ritz\_si\_fc}
    versus \texttt{si\_fc} and \texttt{si\_qwc} (and \texttt{si\_fc} versus
    \texttt{si\_qwc}) without forcing FC to beat QWC.
  \item \textbf{P2:} On stretched ten-qubit LiH, both \texttt{si\_k2} and
    \texttt{ritz\_si\_k2} must achieve \(\ge 12/16\) true precision stops
    (\texttt{target\_met} and \(|E - E_{\mathrm{exact}}| \le 1.6\,\mHa\)),
    \(\le 1/16\) false precision stops, and \(0\) reduced-rank stops.
  \item \textbf{P3:} Under both pure-2Q noise regimes across all three
    eight-qubit banks (6 cells), \texttt{si\_fc\_galic\_tight} must achieve
    strictly smaller exact \(|E_\infty^{\mathrm{noisy}} - E_{\mathrm{exact}}|\)
    and finite-shot RMSE than \texttt{si\_fc\_unconstrained}, and strictly
    fewer false stops whenever unconstrained FC has \(|\mathrm{bias}| >
    1.6\,\mHa\); \texttt{sc\_card\_full} is reported separately as a full-noise
    diagnostic.
\end{itemize}
Shot-ratio intervals use \wpBootAnalysisReps{} independent within-arm
bootstrap resamples (seed \texttt{\wpBootAnalysisSeed}); despite comparisons
being matched by bank, the replicate streams are independent, not paired
observations. The frozen analysis additionally records the fraction \(\hat b\)
of resampled mean ratios at least one and applies Holm--Bonferroni
step-down \cite{holm1979simple} to that diagnostic. These fractions are not
calibrated null-test \(p\)-values (zero means no such resample), so the flag
is reported descriptively and is not the P1a decision rule. Degenerate
bootstrap intervals can reflect a coarse look grid, not zero underlying
uncertainty.

Strict unconditional coverage counts \emph{only} eligible \texttt{target\_met}
stops in the numerator over all declared attempts in the denominator
(exhaustions and failures count as uncovered), with Wilson intervals
\cite{wilson1927probable} and stop-conditional coverage reported separately.
Non-rejection of undercoverage with 16 or 32 outer replicas is not a
validation of nominal coverage at arbitrary sample sizes. A ``false
precision stop'' means an absolute terminal error above the target SE;
even an unbiased Gaussian estimate with SE exactly equal to the target can
exceed that error threshold about 32\% of the time. It is not intrinsically a
controller fault or a confidence-interval miss.

\section{Results}
\label{sec:results}

Execution of the frozen declaration produced
\path{docs/restart/weighted_precision_results.json.gz} containing all
\wpTotalAcq{} acquisitions, \wpTotalShotsExact{} simulated physical shots
(including \wpPilotShotsExact{} paid pilot shots), \wpMeasuredSolves{} measured
solves, and \wpBootSolves{} bootstrap solves. No acquisition failed
(\wpFailed{}/\wpTotalAcq{}); \wpTargetMet{}/\wpTotalAcq{} met their stop
guards and \wpExhausted{}/\wpTotalAcq{} exhausted their schedule (all four in
the opt-in \texttt{beh2\_8q} \texttt{si\_fc\_width} arm).  The independent
checker verified every row, all four occupation-space anchors, all
\wpNoisyAnchorCount{} independent infinite-shot noisy anchors, three full spot
replays, and the declaration-before-record ordering of the version history.

\subsection{P1a: Weighted SI versus unweighted LDF and setting-count paradox}
\label{sec:results_p1a}

\begin{figure*}[t]
  \centering
  \includegraphics[width=\textwidth]{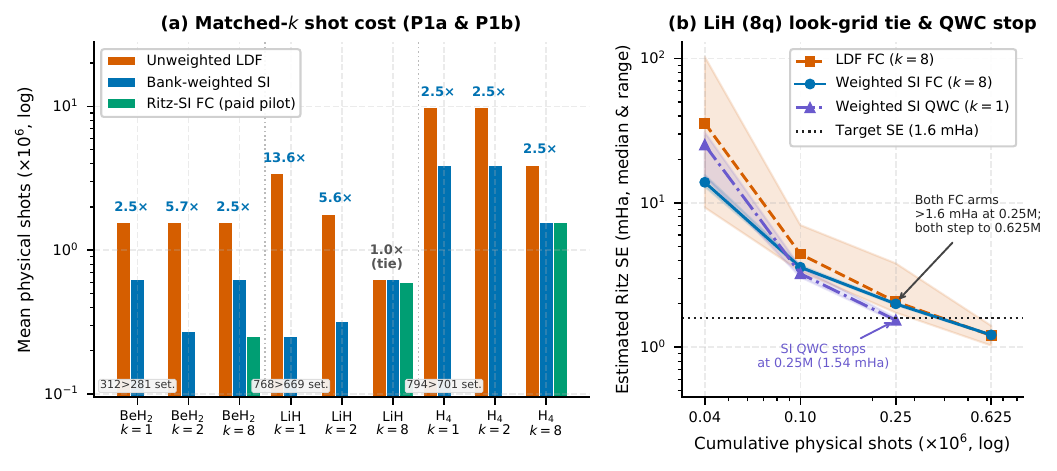}
  \caption{(a)~Mean total physical shots across all nine predeclared P1a pairs
  (\(k \in \{1,2,8\}\)) plus paid-pilot Ritz-SI FC on the three eight-qubit
  molecular banks.  Speedup labels above each pair report unweighted LDF
  divided by weighted SI; gray badges at the bottom of the \(k=1\) pairs note
  that weighted SI uses \emph{more} QWC settings than LDF yet requires
  \(2.5\times\)--\wpPoneaSpeedupMax{}\(\times\) fewer shots.  (b)~Look-by-look
  median estimated Ritz SE (with min--max envelope over 32 replicas) on
  LiH~(8q), showing why both FC arms tie at \(0.625\times 10^6\) shots while
  weighted SI QWC stops early at \(0.25\times 10^6\).}
  \label{fig:shot_savings}
\end{figure*}

\begin{table*}[t]
\centering
\caption{Predeclared P1a matched-\(k\) decision-rule verdicts (A) and P1b
descriptive comparisons (B). Shot ratios are treatment/comparator; 95\%
percentile intervals use \wpBootAnalysisReps{} within-arm resamples (seed
\texttt{\wpBootAnalysisSeed}). Panel A reports the bootstrap tail fraction
\(\hat b\) and descriptive Holm flag, not a calibrated hypothesis-test
\(p\)-value. The coverage lower-tail test uses \(\alpha=0.05/9\). RMSE is in
mHa.}
\label{tab:p1_verdicts}
\compactcolumns
\scriptsize
\begin{tabular}{l c l c c c c l}
\toprule
Molecule & \(k\) & Treatment / comparator & Median (95\% CI) & Mean (95\% CI) & \(\hat b\) (Holm) / RMSE & 95\% boot (\(p_{\mathrm{lo}}\)) & Verdict / \(R/\Rhat\) \\
\midrule
\tableinput{tables/p1_verdicts.tex}
\bottomrule
\end{tabular}
\end{table*}

Weighted sorted insertion passes \wpPoneaPassCount{}/\wpPoneaTotalCount{} P1a
cells (Table~\ref{tab:p1_verdicts}, Panel~A; Fig.~\ref{fig:shot_savings}(a)).
In each passing cell no bootstrap mean-ratio resample reaches one, and
median shots fall to \wpPoneaMedMin{}\(\times\)--\wpPoneaMedMax{}\(\times\)
(\wpPoneaSpeedupMin{}\(\times\)--\wpPoneaSpeedupMax{}\(\times\) fewer mean
shots). These cells achieve \(32/32\) \texttt{target\_met} stops with
\(0\) reduced-rank stops and \(30/32\)--\(32/32\) unconditional nominal
95\% bootstrap-spread coverage.

The QWC (\(k=1\)) cells provide counterexamples to using setting count alone
as a cost proxy: bank-weighted \texttt{si\_qwc} produces \emph{more} settings
than \texttt{ldf\_qwc} on every molecule (Table~\ref{tab:protocol_banks}), yet
requires fewer shots (Fig.~\ref{fig:shot_savings}). Its static bank
\(\Rhat\) is higher, consistent with concentrating important coefficients
into jointly measured groups. This is an improvement in the ideal-shot
objective, not necessarily in wall time or hardware execution cost.

Because P1a requires all nine cells to pass, the conjunction evaluates to
\texttt{all\_passed = false} due to one discrete look-grid tie on LiH~(8q) at
\(k=8\) (\texttt{si\_fc} vs.\ \texttt{ldf\_fc}, ratio \(1.000\)).
Figure~\ref{fig:shot_savings}(b) shows the underlying trajectory: at look~0
(\(0.04\times 10^6\) shots), \texttt{si\_fc} already reduces median SE by
\(2.6\times\) (\wpLihSiFcLookZeroSE{} vs.\ \wpLihLdfFcLookZeroSE{}\,\(\mHa\)),
and at look~2 (\(0.25\times 10^6\) shots) \texttt{si\_fc} exhibits lower and
far tighter SE (\wpLihSiFcLookTwoSE{}\,\(\mHa\), range
[\wpLihSiFcLookTwoLo{}, \wpLihSiFcLookTwoHi{}]) than \texttt{ldf\_fc}
(\wpLihLdfFcLookTwoSE{}\,\(\mHa\), range [\wpLihLdfFcLookTwoLo{},
\wpLihLdfFcLookTwoHi{}]).  Because both lie slightly above the \(1.6\,\mHa\)
stop threshold, the \(2.5\times\) geometric look spacing quantizes all 32
replicas of both arms onto \(0.625\times 10^6\) shots.

\subsection{P1b: Paid-pilot Ritz replanning and state-dependent reversal}
\label{sec:results_p1b}

\begin{figure*}[t]
  \centering
  \includegraphics[width=\textwidth]{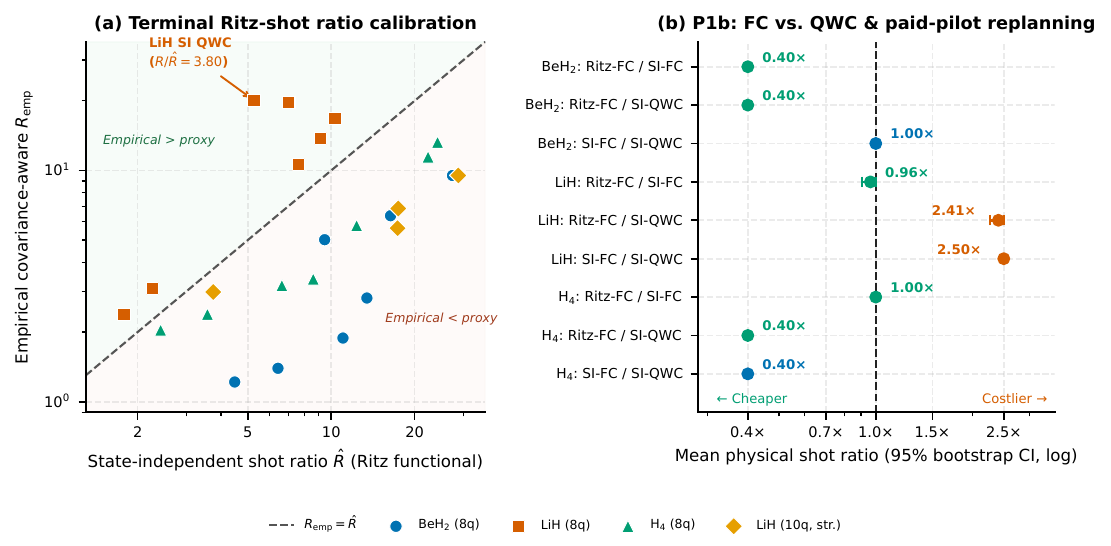}
  \caption{(a)~Median terminal empirical optimal-shot ratio \(\Remp\) versus
  state-independent \(\Rhat\) across Part~A arms (excluding width-guarded
  duplicates). Points above the diagonal have larger empirical ratios than
  predicted by the unit-variance, zero-covariance proxy; this reflects both
  marginal variances and within-setting covariances. (b)~Mean physical-shot ratios (with 95\%
  bootstrap intervals) for the predeclared P1b comparisons. A ratio below
  one favors the numerator. The reversal of FC vs.\ QWC across molecules
  is consistent with the state-dependent variance diagnostics.}
  \label{fig:r_emp_vs_r_hat}
\end{figure*}

Table~\ref{tab:p1_verdicts} (Panel~B) and Fig.~\ref{fig:r_emp_vs_r_hat} show
two distinct effects across molecules:
\begin{enumerate}
  \item \textbf{Paid-pilot Ritz replanning can improve the shot objective.}
    On BeH\(_2\)~(8q), \texttt{ritz\_si\_fc}
    (including its \wpBehTwoRitzPilot{}-shot pilot) cuts total physical shots
    by \(2.5\times\) below \emph{both} \texttt{si\_fc} and \texttt{si\_qwc}
    (\(0.250\times 10^6\) vs.\ \(0.625\times 10^6\), \(32/32\) 95\% coverage,
    RMSE \wpBehTwoRitzRmse{}\,\(\mHa\)).  On LiH~(8q), it trims mean shots
    modestly versus \texttt{si\_fc} (\(0.963\times\)), and on H\(_4\)~(8q) it
    matches \texttt{si\_fc} (\(1.000\times\)). This supports a conditional
    benefit, not a universal advantage of the extra planning stage.
  \item \textbf{FC does not uniformly beat QWC.}  While FC cuts shots by
    \(2.5\times\) relative to \texttt{si\_qwc} on H\(_4\)~(8q), \emph{both} FC
    arms have \(2.5\times\) higher median shot cost than \texttt{si\_qwc} on LiH~(8q)
    (mean costs \(0.625\times 10^6\) / \(0.602\times 10^6\) vs.\ \(0.250\times 10^6\)).
    Figure~\ref{fig:r_emp_vs_r_hat}(a) provides a diagnostic: across all 28
    Part~A cells, \(\Remp/\Rhat\) spans \wpRRatioMin{} to \wpRRatioMax{}.  On
    LiH~(8q), \texttt{si\_qwc} achieves \(\Remp/\Rhat = \wpLihSiQwcRRatio{}\)
    (allowing look-2 SE \wpLihSiQwcLookTwoSE{}\,\(\mHa \le 1.6\,\mHa\)),
    substantially outperforming \texttt{si\_fc} (\wpLihSiFcRRatio{}) and
    \texttt{ritz\_si\_fc} (\wpLihRitzFcRRatio{}). The median terminal
    \(\Rhat\) favors the FC plans, whereas \(\Remp\) favors QWC. This reversal
    is consistent with state-dependent variances and covariances; the terminal
    estimated functionals and adaptive allocation histories differ between
    arms, so it is not a controlled attribution to covariance alone.
\end{enumerate}

\subsection{P2: Stretched ten-qubit rank transfer and width-guard cost}
\label{sec:results_p2}

\begin{figure*}[t]
  \centering
  \includegraphics[width=\textwidth]{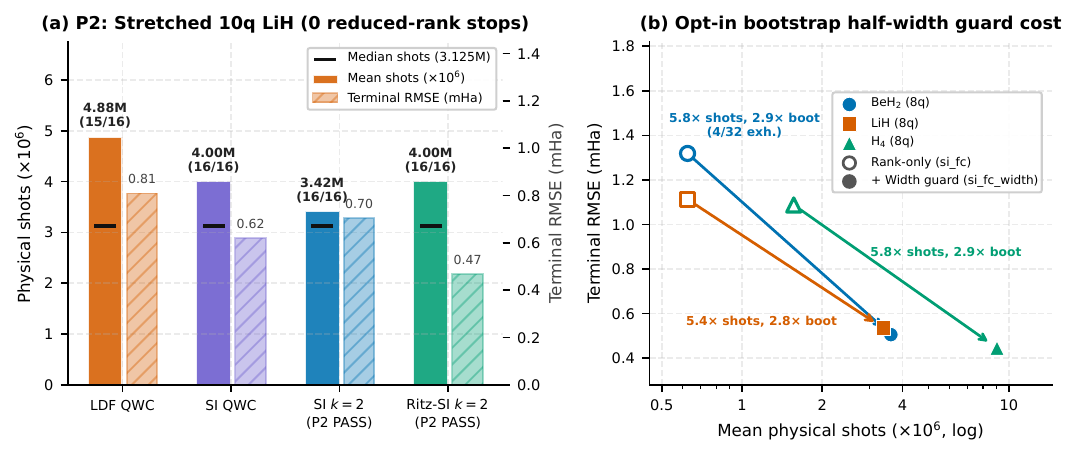}
  \caption{(a)~Predeclared P2 transfer on the stretched ten-qubit LiH bank
  (\wpStretchedWords{} words, rank~4 near \(\tau_S=0.104\)): mean shots
  (solid bars, with true-stop count out of 16), median shots (black ticks), and
  terminal RMSE (hatched bars, right axis).  (b)~Shot-cost vs.\ RMSE shift from
  rank-only \texttt{si\_fc} (open symbols) to opt-in bootstrap half-width
  gating \texttt{si\_fc\_width} (filled symbols) across the three eight-qubit
  banks.}
  \label{fig:stretched_and_width}
\end{figure*}

\begin{table*}[t]
\centering
\caption{Predeclared P2 stretched ten-qubit LiH rank-stress transfer (Panel~A)
and opt-in bootstrap half-width guard overhead on eight-qubit banks (Panel~B).}
\label{tab:p2_and_width}
\compactcolumns
\scriptsize
\begin{tabular}{l c c c c r c l}
\toprule
Arm / Bank & True stops & False (RR) & Med.\ / Mean (M) & Mean ratio (95\% CI) & RMSE (mHa) & 95\% boot (w / cond.) & Verdict / Boot cost \\
\midrule
\tableinput{tables/p2_and_width.tex}
\bottomrule
\end{tabular}
\end{table*}

Both analyses in this subsection evaluate the trade-off between stopping-guard
reliability and physical-shot cost: Panel~A tests whether weighted \(k=2\)
measurement plans preserve rank-margin reliability on a near-cutoff ten-qubit
bank, whereas Panel~B quantifies the extra shot and resampling cost of adding
the opt-in bootstrap half-width guard on the well-conditioned eight-qubit
banks.

Both weighted \(k=2\) arms pass P2 on \texttt{lih\_10q\_stretched}
(Table~\ref{tab:p2_and_width}, Panel~A; Fig.~\ref{fig:stretched_and_width}(a)).
With \(\lambda_{\min} = 0.10417\) sitting just \(1.7\times 10^{-4}\) above
\(\tau_S = 0.104\), \texttt{si\_k2} (\(490\) settings, \(\CZmax=5\), 2Q
depth~1) and \texttt{ritz\_si\_k2} (\(496\)--\(535\) settings, paid pilot
\(249{,}900\) shots) each achieve \(16/16\) true precision stops, \(0/16\)
false stops, and \(0\) reduced-rank stops. All four arms have median cost
\(3.125\times 10^6\) shots on the declared grid, while
\texttt{si\_k2} eliminates high-shot tail runs to reduce mean shots by
\wpStretchedSiKTwoSavePct{}\% versus \texttt{ldf\_qwc}
(\wpStretchedSiKTwoMeanM{}M vs.\ \wpStretchedLdfQwcMeanM{}M) and
\wpStretchedSiKTwoVsSiQwcPct{}\% versus \texttt{si\_qwc}.  Paid-pilot
\texttt{ritz\_si\_k2} reduces mean shots to \wpStretchedRitzRatioLdf{}\(\times\)
\texttt{ldf\_qwc} while achieving the lowest terminal RMSE
(\wpStretchedRitzRmse{} vs.\ \wpStretchedLdfRmse{}\,\(\mHa\)) and narrowest
95\% bootstrap interval (\wpStretchedRitzWidth{}\,\(\mHa\), \(16/16\)
coverage).

Turning from rank-margin reliability to explicit interval-width control,
Table~\ref{tab:p2_and_width} (Panel~B) and
Fig.~\ref{fig:stretched_and_width}(b) quantify the opt-in bootstrap half-width
guard (\texttt{si\_fc\_width} vs.\ \texttt{si\_fc}).  Enforcing a \(1.6\,\mHa\)
six-look adjusted bootstrap half-width lowers terminal RMSE to
\wpWidthRmseMin{}--\wpWidthRmseMax{}\,\(\mHa\) and eliminates almost all
\(1\sigma\) false stops (\(0\)--\(1/32\)), at a cost of
\wpWidthMeanRatioMin{}\(\times\)--\wpWidthMeanRatioMax{}\(\times\) mean
physical shots and
\wpWidthBootRatioMin{}\(\times\)--\wpWidthBootRatioMax{}\(\times\) bootstrap
solves.  On BeH\(_2\)~(8q), \wpExhausted{}/32 attempts exhausted the
\(3.906\times 10^6\)-shot horizon despite actual errors of only
\wpWidthExhErrMin{}--\wpWidthExhErrMax{}\,\(\mHa\) because the Bonferroni-scaled
32-resample bootstrap half-width marginally exceeded \(1.6\,\mHa\)
(stop-conditional 95\% coverage \(28/28\), unconditional \(28/32\)).  This
motivates retaining bootstrap width gating as an opt-in guard. The
comparison quantifies its cost for these banks; it does not establish an
optimal stopping policy.

\subsection{P3: Gate-level noisy readout bias versus variance}
\label{sec:results_p3}

\begin{figure*}[t]
  \centering
  \includegraphics[width=\textwidth]{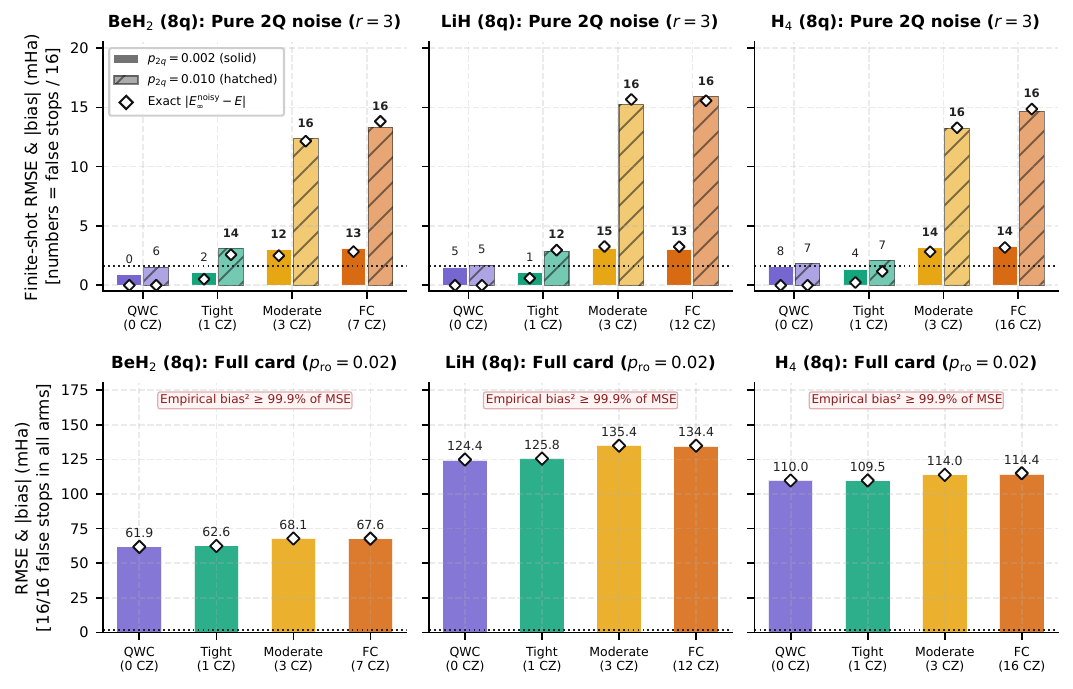}
  \caption{Part~B gate-level noisy readout evaluation across all 36 cells.
  Top row: pure two-qubit depolarizing noise (\(r_{\mathrm{route}}=3\)) at
  \(p_{2\mathrm{q}}=0.002\) (solid bars) and \(p_{2\mathrm{q}}=0.010\) (hatched
  bars); white diamonds mark exact infinite-shot bias
  \(|E_\infty^{\mathrm{noisy}} - E_{\mathrm{exact}}|\), numbers above bars give
  false precision stops out of 16, and dotted lines mark the \(1.6\,\mHa\)
  target. Bottom row: full illustrative superconducting noise model adding
  one-qubit and unmitigated readout bit-flip noise
  (\(p_{\mathrm{ro}}=0.02\)); the squared empirical bias accounts for at least
  99.9\% of MSE in every arm, including zero-2Q QWC.}
  \label{fig:noisy_bias}
\end{figure*}

All \wpPartBAcq{} Part~B acquisitions reached \texttt{target\_met}, despite
often-large differences between the ideal target and the infinite-shot noisy
energy. A precision criterion based on measured sampling variance can be
satisfied around a biased energy; the variance itself can also change under
noise. The rank guard is not a noise-bias detector.

Tight GALIC noise gating (\texttt{si\_fc\_galic\_tight}, \(\CZmax = 1\)) passes
all \wpPthreePassCount{}/\wpPthreeTotalCount{} pure-2Q P3 cells
(Fig.~\ref{fig:noisy_bias}, top row; full 36-cell ledger in
Table~\ref{tab:part_b_noisy}):
\begin{itemize}
  \item At \(p_{2\mathrm{q}}=0.002, r_{\mathrm{route}}=3\), unconstrained FC
    (\(\CZmax = 7\)--\(16\)) suffers \(+2.84\) to \(+3.25\,\mHa\) exact bias
    (\(82\)--\(88\%\) bias\(^2\) share) and
    \wpPtwoLowUncFalseMin{}--\wpPtwoLowUncFalseMax{}\(/16\) false precision
    stops.  Tight GALIC cuts exact bias by \(5.4\times\)--\wpBiasRedMax{}\(\times\)
    (to absolute bias \(0.24\)--\(0.60\,\mHa\)), reduces RMSE by
    \wpRmseRedMin{}\(\times\)--\(3.0\times\), drops false stops to
    \wpPtwoLowTightFalseMin{}--\wpPtwoLowTightFalseMax{}\(/16\), and restores
    95\% unconditional bootstrap coverage to \(16/16\) on all three molecules.
    By contrast, moderate gating (\(\CZmax=3\)) already incurs \(+2.50\) to
    \(+3.27\,\mHa\) bias (\(12\)--\(15/16\) false stops), showing that
    \(\CZmax=3\) is too permissive for \(1.6\,\mHa\) precision at this error
    rate for the tested FC partitions.
  \item At \(p_{2\mathrm{q}}=0.010, r_{\mathrm{route}}=3\), unconstrained FC
    collapses completely (\wpPtwoHighUncBiasMin{}--\wpPtwoHighUncBiasMax{}\,\(\mHa\)
    bias, \(16/16\) false stops, \(0/16\) coverage).  Tight GALIC slashes bias
    and RMSE by \wpBiasRedMin{}\(\times\)--\(12.7\times\) and
    \(4.3\times\)--\wpRmseRedMax{}\(\times\), respectively, and strictly lowers
    false stops on every molecule, but even one logical CZ (three
    routed-equivalent gates) leaves
    \wpPtwoHighTightBiasMin{}--\wpPtwoHighTightBiasMax{}\,\(\mHa\) absolute bias
    (\(7\)--\(14/16\) false stops), leaving zero-2Q \texttt{si\_qwc} as the
    only zero-2Q choice among the tested arms. QWC's exact bias vanishes
    under the pure-2Q model, but finite-shot errors still occur.
  \item Under \texttt{sc\_card\_full} (\(p_{\mathrm{ro}}=0.02\);
    Fig.~\ref{fig:noisy_bias}, bottom row), the combined noise model produces
    \(+\wpScFullBiasMin{}\) to \(+\wpScFullBiasMax{}\,\mHa\) exact bias
    (\(\ge 99.9\%\) bias\(^2\) share, \(16/16\) false stops, \(0/16\) coverage)
    across \emph{every} arm, including \texttt{si\_qwc}. The large residual
    bias in QWC despite having no CZ gates is consistent with readout error
    dominating in this illustrative regime, but the full model also adds
    one-qubit error, so their separate contributions are not identified.
    Two-qubit gating cannot substitute for dedicated readout calibration or
    mitigation \cite{bravyi2021mitigating}; neither is evaluated here.
\end{itemize}

P3 compares tight gating against unconstrained FC, not against QWC. It does
not impose a shot-cost criterion or require absolute error below the target.
The Part~B ledger therefore also reports mean shot cost: for H\(_4\), tight
gating uses \(2.5\times\) the mean shots of unconstrained FC under either
pure-2Q regime, while markedly reducing bias. For LiH, QWC is cheaper than
both FC alternatives and exactly unbiased under pure-2Q noise. Structurally,
tight gating (\(\CZmax=1\)) expands the FC partitions back to \(307\),
\(720\), and \(745\) settings on BeH\(_2\), LiH, and H\(_4\) --- nearly
identical to QWC's \(312\), \(768\), and \(794\) settings
(Table~\ref{tab:part_b_noisy}) --- effectively reverting almost to QWC under
the declared gate cap. By contrast, Part~A's depth-1 disjoint-pair grouping
(\texttt{si\_k2}) achieves much stronger compression (\(44\), \(298\), and
\(399\) settings, \(\CZmax=4\)) while passing the ten-qubit rank-stress test.
Because Part~B was frozen prior to evaluation without a \texttt{si\_k2} arm,
benchmarking depth-1 \(k=2\) grouping directly against tight and moderate
GALIC gating under pure-2Q noise is a natural follow-up. Overall, these
results are accuracy--cost trade-offs, not evidence that tight gating dominates
every alternative.

\section{Scope, Limitations, and Practical Guidance}
\label{sec:discussion}

Our findings yield a concrete engineering recipe for finite-shot subspace
measurement in \pkg{}:
\begin{enumerate}
  \item \textbf{Use variance-weighted sorted insertion as a candidate default.}
    It matches or beats LDF in the nine tested matched-\(k\) cells and
    satisfies the declared reliability and shot-cost criteria in eight.
    Maximum logical 2Q depth is unchanged in these matched plans, but
    setting count and total gate count can increase.
  \item \textbf{Always report empirical \(\Remp\) alongside \(\Rhat\), and test
    \(k=1\) and \(k=2\) before committing to FC.} State-dependent variances
    and covariances can overturn the proxy's ordering, and \(k=2\) depth-1
    readouts meet the rank-stress criterion on the tested ten-qubit bank.
  \item \textbf{Consider paid-pilot Ritz replanning when static weights may
    misalign with the energy sensitivity,} comparing against SI and billing
    every pilot shot to the total budget.
  \item \textbf{Pair GALIC two-qubit noise gating with dedicated readout error
    mitigation.}  At \(p_{2\mathrm{q}}=0.002, r_{\mathrm{route}}=3\),
    \(\CZmax=1\) gives sub-mHa absolute bias on these banks; for the higher
    pure-2Q noise rate, compare against QWC (\(\CZmax=0\)) before accepting
    entangling readouts. The \texttt{sc\_card\_full} regime was included
    specifically to bound the unmitigated systematic-error floor of the full
    device card and demonstrate why any hardware deployment must pair GALIC
    two-qubit gating with calibrated readout error mitigation and a separate
    systematic-error budget.
  \item \textbf{Keep rank/margin confirmation as the default stopping guard and
    bootstrap half-width gating as opt-in} when tighter empirical error
    control justifies its observed shot overhead.
\end{enumerate}

\emph{Limitations.} All experiments use ideal statevector or
explicit depolarizing/bit-flip gate-level simulation on fixed small molecular
banks (8--10 qubits, 3--4 basis directions).  Targets are the declared
fixed-bank projected energies rather than full-CI chemical accuracy.  We do not
claim hardware performance, asymptotic scaling, quantum advantage, or
finite-sample/anytime-valid interval guarantees.  The single failed P1a cell on
LiH (\(k=8\)) and the combined-noise failure of all arms under
\texttt{sc\_card\_full} are retained as first-class experimental outcomes.
All four banks share a fixed two-determinant reference state
\(\cos(0.3)\ket{a}+e^{0.2i}\sin(0.3)\ket{b}\) and small subspace dimension;
because the observed variance savings and the QWC-versus-FC crossover on LiH
depend on state-specific marginal variances and within-setting covariances,
relative performance may shift for more strongly correlated reference states
or larger, more ill-conditioned subspaces (\(m \ge 10\)).
The bank and schedule design and decision rules were informed by development
simulations before freezing. Independent evaluation
replicates test sampling variability conditional on those choices, not
generalization over a population of molecules or devices. The narrow
near-cutoff experiment is not a sweep over ranks or condition numbers, and
the small replicate counts limit conclusions about rare failures. In
particular, \(16/16\) successes corresponds to a two-sided 95\% Wilson
lower bound of about 0.81, not guaranteed success. Alternative adaptive
allocations, noisy \(k=2\) benchmarks, overlapping measurement groups, finer
look grids, and hardware-calibrated mitigation remain relevant comparisons for
future work.

\section{Conclusion}
\label{sec:conclusion}

Within the declared fixed molecular banks, weighted measurement planning,
rank-aware stopping, and gate-level noise limits are complementary rather
than interchangeable controls. Weighted SI improves physical-shot cost in
most matched comparisons, while paid-pilot replanning and wider commuting
groups offer molecule-dependent benefits. The rank-stress experiment
supports transfer to one engineered near-cutoff bank; the noisy experiment
shows why even a reliable sampling-precision stop must be interpreted
separately from systematic energy accuracy. A practical next step is to
prespecify a joint accuracy--cost criterion with independently calibrated
bias mitigation and test it on larger banks and measured hardware data.

\section{Reproducibility and Provenance}
\label{sec:reproducibility}

The study enforces a strict two-stage separation in version control: the
result-free declaration, runner, checker, and analysis code were merged
before any evaluation, and the compressed evaluation artifact
\path{docs/restart/weighted_precision_results.json.gz} was recorded
separately afterwards, following clean-worktree execution under
Python~\wpPythonVer, NumPy~\wpNumpyVer, SciPy~\wpScipyVer, and
single-threaded BLAS.  Routine CI
runs \path{benchmarks/check_weighted_precision.py} with
\texttt{--verify-history} under
that exact declared runtime, re-auditing the declaration-before-record
ordering of the version history, the source digests, all
\wpNoisyAnchorCount{} independent noisy anchors, and all \wpTotalAcq{}
acquisition rows.  Every table, macro, and figure in this manuscript is
generated programmatically from those frozen records and audited by
\path{paper_weighted_precision/check_manuscript.py}.
The frozen declaration, evaluation record, input files, and commands are
available in the public repository at
\url{https://github.com/gutama/clifford_qc}.
The manuscript changes only presentation and analysis of that immutable
record; they do not change any declared endpoint or rerun evaluation seeds.

\appendix
\section{Complete Experimental Ledgers}
\label{app:ledgers}

Tables~\ref{tab:part_a_full} and~\ref{tab:part_b_noisy} report the complete
28-cell Part~A and 36-cell Part~B outcomes underlying Figs.~\ref{fig:shot_savings}--\ref{fig:noisy_bias}
and Tables~\ref{tab:p1_verdicts}--\ref{tab:p2_and_width}.

\begin{table*}[p]
\centering
\caption{Part~A ideal-shot ledger across all 28 \texttt{(case, arm)}
cells (\wpPartAAcq{} acquisitions).  Unconditional coverage counts eligible
\texttt{target\_met} stops over all declared attempts; parenthesized widths
are median interval widths in mHa among stops.  Stop-conditional coverage
equals unconditional coverage in every cell except BeH\(_2\)~(8q)
\texttt{si\_fc\_width} (\(28/28\) at 95\%, \(26/28\) at 68\%, with 4 schedule
exhaustions). RMSE is in mHa across all nonfailed attempts. Dev.\ (s) is the
median illustrative timing proxy using declared logical circuit depths,
preparation and readout times, not measured hardware runtime.}
\label{tab:part_a_full}
\compactcolumns
\scriptsize
\resizebox{\textwidth}{!}{%
\begin{tabular}{l l c c c r r c c c c r}
\toprule
Bank & Arm & Stops (exh.) & T/F & Med./Mean (M) & Dev.\ (s) & RMSE & 95\% boot (w) & 95\% delta (w) & 68\% boot (w) & 68\% delta (w) & \(\Remp/\Rhat\) \\
\midrule
\tableinput{tables/part_a_full.tex}
\bottomrule
\end{tabular}}
\end{table*}

\begin{table*}[p]
\centering
\caption{Complete Part~B gate-level noisy readout ledger across all 36
\texttt{(case, regime, arm)} cells (\wpPartBAcq{} acquisitions; all 576 reached
\texttt{target\_met} with \(0\) reduced-rank stops).  Columns show setting
count (\(\CZmax\)), true/false \(1.6\,\mHa\) stops, mean physical shots
(millions), exact infinite-shot noisy
bias \(E_\infty^{\mathrm{noisy}} - E_{\mathrm{exact}}\), finite-shot empirical
bias \(\pm\) std, RMSE, bias\(^2\) share of MSE, unconditional 95\% bootstrap
coverage, and P3 verdict. Energies, biases, standard deviations and RMSE are
in mHa. The bias-squared share uses finite-shot empirical bias, not the
infinite-shot anchor or a noise-channel attribution.}
\label{tab:part_b_noisy}
\compactcolumns
\scriptsize
\resizebox{\textwidth}{!}{%
\begin{tabular}{l l l c c r r c r r c l}
\toprule
Bank & Regime & Arm & Set.\ (\(\CZmax\)) & T/F & Mean (M) & \(E_\infty\) bias & Emp. bias \(\pm\) std & RMSE & Bias\(^2\) & 95\% boot & P3 \\
\midrule
\tableinput{tables/part_b_noisy.tex}
\bottomrule
\end{tabular}}
\end{table*}

\clearpage

\bibliographystyle{apsrev4-2}
\bibliography{references}

\begin{thebibliography}{21}%
\makeatletter
\providecommand \@ifxundefined [1]{%
 \@ifx{#1\undefined}
}%
\providecommand \@ifnum [1]{%
 \ifnum #1\expandafter \@firstoftwo
 \else \expandafter \@secondoftwo
 \fi
}%
\providecommand \@ifx [1]{%
 \ifx #1\expandafter \@firstoftwo
 \else \expandafter \@secondoftwo
 \fi
}%
\providecommand \natexlab [1]{#1}%
\providecommand \enquote  [1]{``#1''}%
\providecommand \bibnamefont  [1]{#1}%
\providecommand \bibfnamefont [1]{#1}%
\providecommand \citenamefont [1]{#1}%
\providecommand \href@noop [0]{\@secondoftwo}%
\providecommand \href [0]{\begingroup \@sanitize@url \@href}%
\providecommand \@href[1]{\@@startlink{#1}\@@href}%
\providecommand \@@href[1]{\endgroup#1\@@endlink}%
\providecommand \@sanitize@url [0]{\catcode `\\12\catcode `\$12\catcode
  `\&12\catcode `\#12\catcode `\^12\catcode `\_12\catcode `\%12\relax}%
\providecommand \@@startlink[1]{}%
\providecommand \@@endlink[0]{}%
\providecommand \url  [0]{\begingroup\@sanitize@url \@url }%
\providecommand \@url [1]{\endgroup\@href {#1}{\urlprefix }}%
\providecommand \urlprefix  [0]{URL }%
\providecommand \Eprint [0]{\href }%
\providecommand \doibase [0]{https://doi.org/}%
\providecommand \selectlanguage [0]{\@gobble}%
\providecommand \bibinfo  [0]{\@secondoftwo}%
\providecommand \bibfield  [0]{\@secondoftwo}%
\providecommand \translation [1]{[#1]}%
\providecommand \BibitemOpen [0]{}%
\providecommand \bibitemStop [0]{}%
\providecommand \bibitemNoStop [0]{.\EOS\space}%
\providecommand \EOS [0]{\spacefactor3000\relax}%
\providecommand \BibitemShut  [1]{\csname bibitem#1\endcsname}%
\let\auto@bib@innerbib\@empty
\bibitem [{\citenamefont {McClean}\ \emph {et~al.}(2017)\citenamefont
  {McClean}, \citenamefont {Kimchi-Schwartz}, \citenamefont {Carter},\ and\
  \citenamefont {de~Jong}}]{mcclean2017qse}%
  \BibitemOpen
  \bibfield  {author} {\bibinfo {author} {\bibfnamefont {J.~R.}\ \bibnamefont
  {McClean}}, \bibinfo {author} {\bibfnamefont {M.~E.}\ \bibnamefont
  {Kimchi-Schwartz}}, \bibinfo {author} {\bibfnamefont {J.}~\bibnamefont
  {Carter}},\ and\ \bibinfo {author} {\bibfnamefont {W.~A.}\ \bibnamefont
  {de~Jong}},\ }\href {https://doi.org/10.1103/PhysRevA.95.042308} {\bibfield
  {journal} {\bibinfo  {journal} {Physical Review A}\ }\textbf {\bibinfo
  {volume} {95}},\ \bibinfo {pages} {042308} (\bibinfo {year}
  {2017})}\BibitemShut {NoStop}%
\bibitem [{\citenamefont {Huggins}\ \emph {et~al.}(2020)\citenamefont
  {Huggins}, \citenamefont {Lee}, \citenamefont {Baek}, \citenamefont
  {O'Gorman},\ and\ \citenamefont {Whaley}}]{huggins2020nonorthogonal}%
  \BibitemOpen
  \bibfield  {author} {\bibinfo {author} {\bibfnamefont {W.~J.}\ \bibnamefont
  {Huggins}}, \bibinfo {author} {\bibfnamefont {J.}~\bibnamefont {Lee}},
  \bibinfo {author} {\bibfnamefont {U.}~\bibnamefont {Baek}}, \bibinfo {author}
  {\bibfnamefont {B.}~\bibnamefont {O'Gorman}},\ and\ \bibinfo {author}
  {\bibfnamefont {K.~B.}\ \bibnamefont {Whaley}},\ }\href
  {https://doi.org/10.1088/1367-2630/ab867b} {\bibfield  {journal} {\bibinfo
  {journal} {New Journal of Physics}\ }\textbf {\bibinfo {volume} {22}},\
  \bibinfo {pages} {073009} (\bibinfo {year} {2020})}\BibitemShut {NoStop}%
\bibitem [{\citenamefont {Stair}\ \emph {et~al.}(2020)\citenamefont {Stair},
  \citenamefont {Huang},\ and\ \citenamefont {Evangelista}}]{stair2020krylov}%
  \BibitemOpen
  \bibfield  {author} {\bibinfo {author} {\bibfnamefont {N.~H.}\ \bibnamefont
  {Stair}}, \bibinfo {author} {\bibfnamefont {R.}~\bibnamefont {Huang}},\ and\
  \bibinfo {author} {\bibfnamefont {F.~A.}\ \bibnamefont {Evangelista}},\
  }\href {https://doi.org/10.1021/acs.jctc.9b01125} {\bibfield  {journal}
  {\bibinfo  {journal} {Journal of Chemical Theory and Computation}\ }\textbf
  {\bibinfo {volume} {16}},\ \bibinfo {pages} {2236} (\bibinfo {year}
  {2020})}\BibitemShut {NoStop}%
\bibitem [{\citenamefont {Utama}\ and\ \citenamefont
  {Dipojono}(2026{\natexlab{a}})}]{utama2026operatorcentric}%
  \BibitemOpen
  \bibfield  {author} {\bibinfo {author} {\bibfnamefont {G.}~\bibnamefont
  {Utama}}\ and\ \bibinfo {author} {\bibfnamefont {H.~K.}\ \bibnamefont
  {Dipojono}},\ }\href {https://doi.org/10.48550/arXiv.2607.17443} {\bibinfo
  {title} {Operator-centric {C}lifford algebra for variational eigensolvers and
  finite-shot adaptive selection}} (\bibinfo {year} {2026}{\natexlab{a}}),\
  \Eprint {https://arxiv.org/abs/2607.17443} {arXiv:2607.17443 [quant-ph]}
  \BibitemShut {NoStop}%
\bibitem [{\citenamefont {Utama}\ and\ \citenamefont
  {Dipojono}(2026{\natexlab{b}})}]{utama2026acase}%
  \BibitemOpen
  \bibfield  {author} {\bibinfo {author} {\bibfnamefont {G.}~\bibnamefont
  {Utama}}\ and\ \bibinfo {author} {\bibfnamefont {H.~K.}\ \bibnamefont
  {Dipojono}},\ }\href {https://doi.org/10.48550/arXiv.2608.00560} {\bibinfo
  {title} {Adaptive operator-generated subspaces for effective many-body
  {H}amiltonians}} (\bibinfo {year} {2026}{\natexlab{b}}),\ \Eprint
  {https://arxiv.org/abs/2608.00560} {arXiv:2608.00560 [quant-ph]} \BibitemShut
  {NoStop}%
\bibitem [{\citenamefont {Epperly}\ \emph {et~al.}(2022)\citenamefont
  {Epperly}, \citenamefont {Lin},\ and\ \citenamefont
  {Nakatsukasa}}]{epperly2022theory}%
  \BibitemOpen
  \bibfield  {author} {\bibinfo {author} {\bibfnamefont {E.~N.}\ \bibnamefont
  {Epperly}}, \bibinfo {author} {\bibfnamefont {L.}~\bibnamefont {Lin}},\ and\
  \bibinfo {author} {\bibfnamefont {Y.}~\bibnamefont {Nakatsukasa}},\ }\href
  {https://doi.org/10.1137/21M145954X} {\bibfield  {journal} {\bibinfo
  {journal} {SIAM Journal on Matrix Analysis and Applications}\ }\textbf
  {\bibinfo {volume} {43}},\ \bibinfo {pages} {1263} (\bibinfo {year}
  {2022})}\BibitemShut {NoStop}%
\bibitem [{\citenamefont {Lee}\ \emph {et~al.}(2024)\citenamefont {Lee},
  \citenamefont {Lee},\ and\ \citenamefont {Huh}}]{lee2024sampling}%
  \BibitemOpen
  \bibfield  {author} {\bibinfo {author} {\bibfnamefont {G.}~\bibnamefont
  {Lee}}, \bibinfo {author} {\bibfnamefont {D.}~\bibnamefont {Lee}},\ and\
  \bibinfo {author} {\bibfnamefont {J.}~\bibnamefont {Huh}},\ }\href
  {https://doi.org/10.22331/q-2024-09-19-1477} {\bibfield  {journal} {\bibinfo
  {journal} {Quantum}\ }\textbf {\bibinfo {volume} {8}},\ \bibinfo {pages}
  {1477} (\bibinfo {year} {2024})}\BibitemShut {NoStop}%
\bibitem [{\citenamefont {Zhang}\ \emph {et~al.}(2024)\citenamefont {Zhang},
  \citenamefont {Wang}, \citenamefont {Xu},\ and\ \citenamefont
  {Li}}]{zhang2024measurement}%
  \BibitemOpen
  \bibfield  {author} {\bibinfo {author} {\bibfnamefont {Z.}~\bibnamefont
  {Zhang}}, \bibinfo {author} {\bibfnamefont {A.}~\bibnamefont {Wang}},
  \bibinfo {author} {\bibfnamefont {X.}~\bibnamefont {Xu}},\ and\ \bibinfo
  {author} {\bibfnamefont {Y.}~\bibnamefont {Li}},\ }\href
  {https://doi.org/10.22331/q-2024-08-13-1438} {\bibfield  {journal} {\bibinfo
  {journal} {Quantum}\ }\textbf {\bibinfo {volume} {8}},\ \bibinfo {pages}
  {1438} (\bibinfo {year} {2024})}\BibitemShut {NoStop}%
\bibitem [{\citenamefont {Verteletskyi}\ \emph {et~al.}(2020)\citenamefont
  {Verteletskyi}, \citenamefont {Yen},\ and\ \citenamefont
  {Izmaylov}}]{verteletskyi2020grouping}%
  \BibitemOpen
  \bibfield  {author} {\bibinfo {author} {\bibfnamefont {V.}~\bibnamefont
  {Verteletskyi}}, \bibinfo {author} {\bibfnamefont {T.-C.}\ \bibnamefont
  {Yen}},\ and\ \bibinfo {author} {\bibfnamefont {A.~F.}\ \bibnamefont
  {Izmaylov}},\ }\href {https://doi.org/10.1063/1.5141458} {\bibfield
  {journal} {\bibinfo  {journal} {Journal of Chemical Physics}\ }\textbf
  {\bibinfo {volume} {152}},\ \bibinfo {pages} {124114} (\bibinfo {year}
  {2020})}\BibitemShut {NoStop}%
\bibitem [{\citenamefont {Izmaylov}\ \emph {et~al.}(2020)\citenamefont
  {Izmaylov}, \citenamefont {Yen}, \citenamefont {Lang},\ and\ \citenamefont
  {Verteletskyi}}]{izmaylov2020unitary}%
  \BibitemOpen
  \bibfield  {author} {\bibinfo {author} {\bibfnamefont {A.~F.}\ \bibnamefont
  {Izmaylov}}, \bibinfo {author} {\bibfnamefont {T.-C.}\ \bibnamefont {Yen}},
  \bibinfo {author} {\bibfnamefont {R.~A.}\ \bibnamefont {Lang}},\ and\
  \bibinfo {author} {\bibfnamefont {V.}~\bibnamefont {Verteletskyi}},\ }\href
  {https://doi.org/10.1021/acs.jctc.9b00791} {\bibfield  {journal} {\bibinfo
  {journal} {Journal of Chemical Theory and Computation}\ }\textbf {\bibinfo
  {volume} {16}},\ \bibinfo {pages} {190} (\bibinfo {year} {2020})}\BibitemShut
  {NoStop}%
\bibitem [{\citenamefont {Crawford}\ \emph {et~al.}(2021)\citenamefont
  {Crawford}, \citenamefont {van Straaten}, \citenamefont {Wang}, \citenamefont
  {Parks}, \citenamefont {Campbell},\ and\ \citenamefont
  {Brierley}}]{crawford2021efficient}%
  \BibitemOpen
  \bibfield  {author} {\bibinfo {author} {\bibfnamefont {O.}~\bibnamefont
  {Crawford}}, \bibinfo {author} {\bibfnamefont {B.}~\bibnamefont {van
  Straaten}}, \bibinfo {author} {\bibfnamefont {D.}~\bibnamefont {Wang}},
  \bibinfo {author} {\bibfnamefont {T.}~\bibnamefont {Parks}}, \bibinfo
  {author} {\bibfnamefont {E.}~\bibnamefont {Campbell}},\ and\ \bibinfo
  {author} {\bibfnamefont {S.}~\bibnamefont {Brierley}},\ }\href
  {https://doi.org/10.22331/q-2021-01-20-385} {\bibfield  {journal} {\bibinfo
  {journal} {Quantum}\ }\textbf {\bibinfo {volume} {5}},\ \bibinfo {pages}
  {385} (\bibinfo {year} {2021})}\BibitemShut {NoStop}%
\bibitem [{\citenamefont {Miller}\ \emph {et~al.}(2024)\citenamefont {Miller},
  \citenamefont {Fischer}, \citenamefont {Levi}, \citenamefont {Kuehnke},
  \citenamefont {Sokolov}, \citenamefont {Barkoutsos}, \citenamefont {Eisert},\
  and\ \citenamefont {Tavernelli}}]{miller2024hardware}%
  \BibitemOpen
  \bibfield  {author} {\bibinfo {author} {\bibfnamefont {D.}~\bibnamefont
  {Miller}}, \bibinfo {author} {\bibfnamefont {L.~E.}\ \bibnamefont {Fischer}},
  \bibinfo {author} {\bibfnamefont {K.}~\bibnamefont {Levi}}, \bibinfo {author}
  {\bibfnamefont {E.~J.}\ \bibnamefont {Kuehnke}}, \bibinfo {author}
  {\bibfnamefont {I.~O.}\ \bibnamefont {Sokolov}}, \bibinfo {author}
  {\bibfnamefont {P.~K.}\ \bibnamefont {Barkoutsos}}, \bibinfo {author}
  {\bibfnamefont {J.}~\bibnamefont {Eisert}},\ and\ \bibinfo {author}
  {\bibfnamefont {I.}~\bibnamefont {Tavernelli}},\ }\href
  {https://doi.org/10.1038/s41534-024-00901-1} {\bibfield  {journal} {\bibinfo
  {journal} {npj Quantum Information}\ }\textbf {\bibinfo {volume} {10}},\
  \bibinfo {pages} {122} (\bibinfo {year} {2024})}\BibitemShut {NoStop}%
\bibitem [{\citenamefont {DalFavero}\ \emph {et~al.}(2025)\citenamefont
  {DalFavero}, \citenamefont {Sarkar}, \citenamefont {Rowland}, \citenamefont
  {Camps}, \citenamefont {Sawaya},\ and\ \citenamefont
  {LaRose}}]{dalfavero2025kcommutativity}%
  \BibitemOpen
  \bibfield  {author} {\bibinfo {author} {\bibfnamefont {B.}~\bibnamefont
  {DalFavero}}, \bibinfo {author} {\bibfnamefont {R.}~\bibnamefont {Sarkar}},
  \bibinfo {author} {\bibfnamefont {J.}~\bibnamefont {Rowland}}, \bibinfo
  {author} {\bibfnamefont {D.}~\bibnamefont {Camps}}, \bibinfo {author}
  {\bibfnamefont {N.~P.~D.}\ \bibnamefont {Sawaya}},\ and\ \bibinfo {author}
  {\bibfnamefont {R.}~\bibnamefont {LaRose}},\ }\href
  {https://doi.org/10.1103/y6hl-gzr4} {\bibfield  {journal} {\bibinfo
  {journal} {Physical Review A}\ }\textbf {\bibinfo {volume} {112}},\ \bibinfo
  {pages} {052407} (\bibinfo {year} {2025})},\ \Eprint
  {https://arxiv.org/abs/2312.11840} {arXiv:2312.11840 [quant-ph]} \BibitemShut
  {NoStop}%
\bibitem [{\citenamefont {Burns}\ \emph {et~al.}(2024)\citenamefont {Burns},
  \citenamefont {Liu}, \citenamefont {Stein}, \citenamefont {Peng},
  \citenamefont {Kowalski},\ and\ \citenamefont {Li}}]{burns2024galic}%
  \BibitemOpen
  \bibfield  {author} {\bibinfo {author} {\bibfnamefont {M.~X.}\ \bibnamefont
  {Burns}}, \bibinfo {author} {\bibfnamefont {C.}~\bibnamefont {Liu}}, \bibinfo
  {author} {\bibfnamefont {S.~A.}\ \bibnamefont {Stein}}, \bibinfo {author}
  {\bibfnamefont {B.}~\bibnamefont {Peng}}, \bibinfo {author} {\bibfnamefont
  {K.}~\bibnamefont {Kowalski}},\ and\ \bibinfo {author} {\bibfnamefont
  {A.}~\bibnamefont {Li}},\ }\href {https://doi.org/10.48550/arXiv.2409.00576}
  {\bibinfo {title} {{GALIC}: Hybrid multi-qubitwise {Pauli} grouping for
  quantum computing measurement}} (\bibinfo {year} {2024}),\ \Eprint
  {https://arxiv.org/abs/2409.00576} {arXiv:2409.00576 [quant-ph]} \BibitemShut
  {NoStop}%
\bibitem [{\citenamefont {Efron}(1979)}]{efron1979bootstrap}%
  \BibitemOpen
  \bibfield  {author} {\bibinfo {author} {\bibfnamefont {B.}~\bibnamefont
  {Efron}},\ }\href {https://doi.org/10.1214/aos/1176344552} {\bibfield
  {journal} {\bibinfo  {journal} {The Annals of Statistics}\ }\textbf {\bibinfo
  {volume} {7}},\ \bibinfo {pages} {1} (\bibinfo {year} {1979})}\BibitemShut
  {NoStop}%
\bibitem [{\citenamefont {Utama}\ and\ \citenamefont
  {Dipojono}(2026{\natexlab{c}})}]{utama2026cliffordqc}%
  \BibitemOpen
  \bibfield  {author} {\bibinfo {author} {\bibfnamefont {G.}~\bibnamefont
  {Utama}}\ and\ \bibinfo {author} {\bibfnamefont {H.~K.}\ \bibnamefont
  {Dipojono}},\ }\href {https://doi.org/10.48550/arXiv.2609.23059} {\bibinfo
  {title} {\texttt{clifford\_qc}: A {P}ython toolkit for quantum simulation}}
  (\bibinfo {year} {2026}{\natexlab{c}}),\ \Eprint
  {https://arxiv.org/abs/2609.23059} {arXiv:2609.23059 [quant-ph]} \BibitemShut
  {NoStop}%
\bibitem [{\citenamefont {Yen}\ \emph {et~al.}(2023)\citenamefont {Yen},
  \citenamefont {Ganeshram},\ and\ \citenamefont
  {Izmaylov}}]{yen2023covariance}%
  \BibitemOpen
  \bibfield  {author} {\bibinfo {author} {\bibfnamefont {T.-C.}\ \bibnamefont
  {Yen}}, \bibinfo {author} {\bibfnamefont {A.}~\bibnamefont {Ganeshram}},\
  and\ \bibinfo {author} {\bibfnamefont {A.~F.}\ \bibnamefont {Izmaylov}},\
  }\href {https://doi.org/10.1038/s41534-023-00683-y} {\bibfield  {journal}
  {\bibinfo  {journal} {npj Quantum Information}\ }\textbf {\bibinfo {volume}
  {9}},\ \bibinfo {pages} {14} (\bibinfo {year} {2023})}\BibitemShut {NoStop}%
\bibitem [{\citenamefont {Knowles}\ and\ \citenamefont
  {Handy}(1989)}]{knowles1989fcidump}%
  \BibitemOpen
  \bibfield  {author} {\bibinfo {author} {\bibfnamefont {P.~J.}\ \bibnamefont
  {Knowles}}\ and\ \bibinfo {author} {\bibfnamefont {N.~C.}\ \bibnamefont
  {Handy}},\ }\href {https://doi.org/10.1016/0010-4655(89)90033-7} {\bibfield
  {journal} {\bibinfo  {journal} {Computer Physics Communications}\ }\textbf
  {\bibinfo {volume} {54}},\ \bibinfo {pages} {75} (\bibinfo {year}
  {1989})}\BibitemShut {NoStop}%
\bibitem [{\citenamefont {Holm}(1979)}]{holm1979simple}%
  \BibitemOpen
  \bibfield  {author} {\bibinfo {author} {\bibfnamefont {S.}~\bibnamefont
  {Holm}},\ }\href@noop {} {\bibfield  {journal} {\bibinfo  {journal}
  {Scandinavian Journal of Statistics}\ }\textbf {\bibinfo {volume} {6}},\
  \bibinfo {pages} {65} (\bibinfo {year} {1979})}\BibitemShut {NoStop}%
\bibitem [{\citenamefont {Wilson}(1927)}]{wilson1927probable}%
  \BibitemOpen
  \bibfield  {author} {\bibinfo {author} {\bibfnamefont {E.~B.}\ \bibnamefont
  {Wilson}},\ }\href {https://doi.org/10.1080/01621459.1927.10502953}
  {\bibfield  {journal} {\bibinfo  {journal} {Journal of the American
  Statistical Association}\ }\textbf {\bibinfo {volume} {22}},\ \bibinfo
  {pages} {209} (\bibinfo {year} {1927})}\BibitemShut {NoStop}%
\bibitem [{\citenamefont {Bravyi}\ \emph {et~al.}(2021)\citenamefont {Bravyi},
  \citenamefont {Sheldon}, \citenamefont {Kandala}, \citenamefont {McKay},\
  and\ \citenamefont {Gambetta}}]{bravyi2021mitigating}%
  \BibitemOpen
  \bibfield  {author} {\bibinfo {author} {\bibfnamefont {S.}~\bibnamefont
  {Bravyi}}, \bibinfo {author} {\bibfnamefont {S.}~\bibnamefont {Sheldon}},
  \bibinfo {author} {\bibfnamefont {A.}~\bibnamefont {Kandala}}, \bibinfo
  {author} {\bibfnamefont {D.~C.}\ \bibnamefont {McKay}},\ and\ \bibinfo
  {author} {\bibfnamefont {J.~M.}\ \bibnamefont {Gambetta}},\ }\href
  {https://doi.org/10.1103/PhysRevA.103.042605} {\bibfield  {journal} {\bibinfo
   {journal} {Physical Review A}\ }\textbf {\bibinfo {volume} {103}},\ \bibinfo
  {pages} {042605} (\bibinfo {year} {2021})}\BibitemShut {NoStop}%
\end{thebibliography}%

\end{document}